\documentclass[aps,prl,showpacs,floatfix,twocolumn,byrevtex,superscriptaddress]{revtex4-1}
\usepackage{amsmath}
\usepackage{amssymb}
\usepackage{amstext}
\usepackage{amsopn}
\usepackage{amsfonts}
\usepackage{amsxtra}
\usepackage[english]{babel}
\usepackage{graphicx}
\usepackage{float}
\usepackage{bm}
\usepackage{multirow}
\usepackage{dcolumn}
\usepackage{color}
\usepackage{hyperref}

\newcommand{\icm}{\ensuremath{\textrm{cm}^{-1}}}

\newcommand{\CVS}{CsV$_{3}$Sb$_{5}$}
\newcommand{\KVS}{KV$_{3}$Sb$_{5}$}
\newcommand{\RVS}{RbV$_{3}$Sb$_{5}$}
\newcommand{\AVS}{$A$V$_{3}$Sb$_{5}$}
\newcommand{\TCDW}{$T_{\text{CDW}}$}

\begin{document}

\title{Origin of the Charge Density Wave in the Kagome Metal CsV$_{3}$Sb$_{5}$ as Revealed by Optical Spectroscopy}
\author{Xiaoxiang Zhou}
\thanks{These authors contributed equally to this work.}
\affiliation{National Laboratory of Solid State Microstructures and Department of Physics, Collaborative Innovation Center of Advanced Microstructures, Nanjing University, Nanjing 210093, China}
\author{Yongkai Li}
\thanks{These authors contributed equally to this work.}
\affiliation{Key Laboratory of Advanced Optoelectronic Quantum Architecture and Measurement, Ministry of Education, School of Physics, Beijing Institute of Technology, Beijing 100081, China}
\affiliation{Micronano Center, Beijing Key Lab of Nanophotonics and Ultrafine Optoelectronic Systems, Beijing Institute of Technology, Beijing 100081, China}
\author{Xinwei Fan}
\thanks{These authors contributed equally to this work.}
\author{Jiahao Hao}
\author{Yaomin Dai}
\email{ymdai@nju.edu.cn}
\affiliation{National Laboratory of Solid State Microstructures and Department of Physics, Collaborative Innovation Center of Advanced Microstructures, Nanjing University, Nanjing 210093, China}
\author{Zhiwei Wang}
\email{zhiweiwang@bit.edu.cn}
\author{Yugui Yao}
\affiliation{Key Laboratory of Advanced Optoelectronic Quantum Architecture and Measurement, Ministry of Education, School of Physics, Beijing Institute of Technology, Beijing 100081, China}
\affiliation{Micronano Center, Beijing Key Lab of Nanophotonics and Ultrafine Optoelectronic Systems, Beijing Institute of Technology, Beijing 100081, China}
\author{Hai-Hu Wen}
\email{hhwen@nju.edu.cn}
\affiliation{National Laboratory of Solid State Microstructures and Department of Physics, Collaborative Innovation Center of Advanced Microstructures, Nanjing University, Nanjing 210093, China}

\date{\today}
%
%

\begin{abstract}
We report on a study of the optical properties of CsV$_{3}$Sb$_{5}$ at a large number of temperatures above and below the charge-density-wave (CDW) transition. Above the CDW transition, the low-frequency optical conductivity reveals two Drude components with distinct widths. An examination of the band structure allows us to ascribe the narrow Drude to a light electron-like and multiple Dirac bands, and the broad Drude to heavy hole bands near the $M$ points which form saddle points near the Fermi level. Upon entering the CDW state, the opening of the CDW gap is clearly observed. A large portion of the broad Drude is removed by the gap, whereas the narrow Drude is not affected. Meanwhile, an absorption peak associated with interband transitions involving the saddle points shifts to higher energy and grows in weight. These observations attest to the importance of saddle point nesting in driving the CDW instability in CsV$_{3}$Sb$_{5}$.
\end{abstract}



\maketitle

%
%
Kagome lattice has long been serving as a fertile ground for exploring exotic physics, because it supports a wide variety of intriguing topological states and electronic instabilities. For example, electrons in a kagome lattice usually form flat bands and symmetry-protected Dirac points, giving rise to various interesting quantum phenomena associated with nontrivial band topology~\cite{Guo2009PRB,Mazin2014NC,Ye2018Nature,Kang2020NM,Liu2020NC,Yin2019NP}. Furthermore, depending on the electron filling, on-site repulsion $U$, and nearest-neighbor Coulomb interaction $V$, the ground state of a kagome lattice system can be a quantum spin liquid~\cite{Balents2010Nature,Yan2011Science}, charge bond order~\cite{Wang2013PRB,Kiesel2013PRL}, superconductor~\cite{Ko2009PRB,Yu2012PRB,Wang2013PRB,Kiesel2013PRL}, charge density wave (CDW)~\cite{Wang2013PRB}, or spin density wave (SDW)~\cite{Yu2012PRB}.

Recently, a new family of kagome metals \AVS\ ($A$ = K, Rb, or Cs) was discovered~\cite{Ortiz2019PRM}, prompting an intense exploration into the exotic physics in these compounds~\cite{Ortiz2020PRL,Ortiz2021PRM,Yin2021CPL,Chen2021arXiv,Chen2021arXiv2,Zhao2021arXiv,Duan2021arXiv,Chen2021arXiv1,Liang2021arXiv,Jiang2020arXiv,Yang2020SA,Yu2021arXiv,Feng2021arXiv,Denner2021arXiv,Li2021arXiv,Tan2021arXiv,Uykur2021arXiv,Zhang2021arXiv,Zhao2021arXiv1,Zhao2021arXiv2}.
Superconductivity with a transition temperature $T_{c}$ of 0.93, 0.92, and 2.5~K has been reported in \KVS, \RVS, and \CVS, respectively~\cite{Ortiz2020PRL,Ortiz2021PRM,Yin2021CPL}. By applying pressure, $T_{c}$ of \CVS\ can be enhanced to about 8~K at 2~GPa~\cite{Chen2021arXiv,Chen2021arXiv2}. While an ultralow-temperature thermal transport study has revealed possible nodal superconductivity in \CVS~\cite{Zhao2021arXiv}, magnetic penetration depth and specific heat measurements have provided evidence for nodeless superconducting gaps~\cite{Duan2021arXiv}. In addition to superconductivity, \AVS\ undergoes a CDW transition at \TCDW\ = 78, 103, and 94~K for $A$ = K, Rb, and Cs, respectively~\cite{Ortiz2019PRM,Ortiz2020PRL,Ortiz2021PRM,Yin2021CPL}. X-ray diffraction~\cite{Ortiz2020PRL} and scanning tunneling microscopy (STM)~\cite{Jiang2020arXiv,Chen2021arXiv1,Liang2021arXiv} measurements have revealed a 2$\times$2 superlattice, attesting to the formation of charge ordering. Moreover, giant anomalous Hall effect~\cite{Yang2020SA,Yu2021arXiv} and multiple topologically protected Dirac bands near the Fermi level ($E_{F}$)~\cite{Ortiz2020PRL,Yin2021CPL} have also been observed in these compounds.

Although some experimental results have demonstrated that the CDW order has strong influence on superconductivity and nontrivial topological bands~\cite{Chen2021arXiv,Yu2021arXiv,Zhao2021arXiv,Jiang2020arXiv}, the driving mechanism of the CDW order remains unclear. Theoretical calculations suggest that the CDW transition is driven by the Peierls instability related to the Fermi surface (FS) nesting and the softening of an acoustic phonon mode~\cite{Tan2021arXiv}, which is supported by STM measurements~\cite{Jiang2020arXiv,Chen2021arXiv1}, but on the other hand, a hard X-ray scattering study on \RVS\ and \CVS\ fails to detect the expected acoustic phonon anomaly at the CDW wave vector, pointing to an unconventional electronic-driven mechanism~\cite{Li2021arXiv}.

In this Letter, we investigate the optical properties of \CVS\ at 15 temperatures above and below \TCDW. Two Drude components (a narrow and a broad one) yield a good description of the low-frequency optical conductivity. Through close inspection of the band structure, we ascribe the narrow Drude to a light electron-like and multiple Dirac bands, and the broad Drude to heavy hole bands near the $M$ points which have saddle points near $E_{F}$. Below \TCDW, the optical conductivity clearly demonstrates the formation of the CDW gap, which significantly suppresses the weight of the broad Drude, while does not have noticeable effect on the narrow Drude. Simultaneously, interband transitions involving states near the saddle points are enhanced and shift to higher energy. These experimental results are in favor of the scenario that nesting of the saddle points at $M$ induces the CDW instability in \CVS.

%
%

\begin{figure*}[tb]
\includegraphics[width=\textwidth]{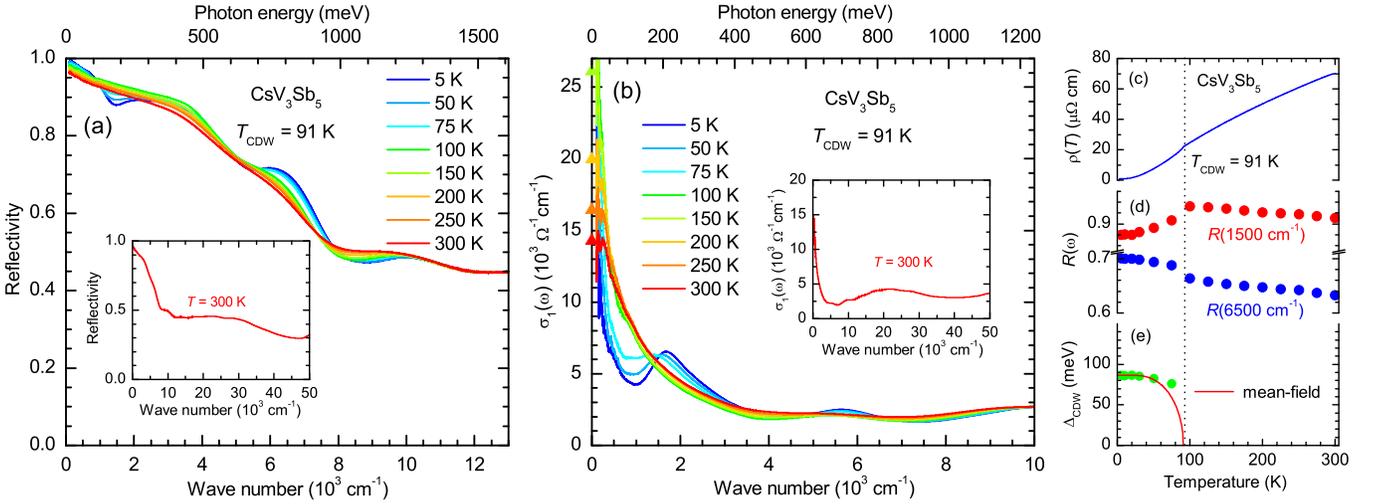}
\caption{(a) and (b) show the reflectivity $R(\omega)$ and the real part of the optical conductivity $\sigma_{1}(\omega)$ of \CVS, respectively, at several representative temperatures above and below $T_{\text{CDW}}$. The insets of (a) and (b) display $R(\omega)$ and $\sigma_{1}(\omega)$ up to 50\,000~\icm\ at 300~K, respectively. The solid triangles on the $y$ axis of (b) denote the dc conductivity at different temperatures obtained from transport measurements. (c) The $T$-dependent resistivity $\rho(T)$ of \CVS. (d) The $T$ dependence of $R(\omega)$ at $\omega = 1500$~\icm\ (red solid circles) and $\omega = 6500$~\icm\ (blue solid circles). (e) The CDW gap $\Delta_{\text{CDW}}$ at different temperatures (green solid circles) and the BCS mean-field behavior (red solid curve).}
\label{CRefS1}
\end{figure*}

Single crystals of \CVS\ were synthesized via a self-flux method~\cite{Ortiz2019PRM}. Figure~\ref{CRefS1}(c) shows the $T$-dependent resistivity $\rho(T)$ of our sample, which exhibits a CDW transition at $T_{\text{CDW}} = 91$~K. The \emph{ab}-plane reflectivity $R(\omega)$ of \CVS\ was measured at a near-normal angle of incidence on a newly cleaved surface using a Bruker Vertex 80V Fourier transform spectrometer equipped with an \emph{in situ} gold evaporation technique~\cite{Homes1993}. Data in the frequency range of 100--12\,000~\icm\ were collected at 15 temperatures from 300 down to 5~K. Then, an AvaSpec-2048$\times$14 optical fiber spectrometer was employed to extend $R(\omega)$ to 50\,000~\icm\ at room temperature. The real part of the optical conductivity $\sigma_{1}(\omega)$ was obtained from a Kramers-Kronig analysis of the measured $R(\omega)$~\cite{Dressel2002}.

Figure~\ref{CRefS1}(a) displays $R(\omega)$ of \CVS\ at several representative temperatures above and below $T_{\text{CDW}}$ up to 13\,000~\icm; the inset shows $R(\omega)$ at 300~K up to 50\,000~\icm. Above \TCDW, $R(\omega)$ approaches unity in the low-frequency limit and increases with decreasing $T$ in the far-infrared range, agreeing well with the metallic nature of this compound. Below \TCDW, for example at 5~K (blue curve), a dramatic suppression of $R(\omega)$ develops at about 1500~\icm\ alongside a hump-like feature at about 6500~\icm. The evolution of $R(1500~\icm)$ and $R(6500~\icm)$ with $T$ is traced out in Fig.~\ref{CRefS1}(d). Both the suppression of $R(1500~\icm)$ and the enhancement of $R(6500~\icm)$ occur at \TCDW, suggesting that they are intimately related to the CDW transition.

More straightforward information can be obtained from $\sigma_{1}(\omega)$, as it is directly linked to the joint density of states~\cite{Basov2005RMP}. Figure~\ref{CRefS1}(b) shows $\sigma_{1}(\omega)$ of \CVS\ at different temperatures above and below \TCDW. The solid triangles on the $y$ axis denote the dc conductivity $\sigma_{dc}$ at different temperatures obtained from transport measurements, which agrees quite well with the zero-frequency extrapolation of $\sigma_{1}(\omega)$. For $T > T_{\text{CDW}}$, the low-frequency $\sigma_{1}(\omega)$ exhibits a pronounced Drude response, i.e. a peak centered at zero frequency, which is a typical characteristic of metals. As $T$ is lowered, the Drude peak narrows, reflecting the reduction of the quasiparticle scattering rate. In the high-frequency range, two broad bands associated with interband transitions can be resolved at about 5500 and 10\,000~\icm. For $T < T_{\text{CDW}}$, $\sigma_{1}(\omega)$ below about 1500~\icm\ is significantly suppressed, and the lost spectral weight [the area under $\sigma_{1}(\omega)$] is transferred to a higher frequency range 1500--4000~\icm, resulting in a conspicuous peak at about 1700~\icm. This gives strong optical evidence for the opening of a density-wave gap~\cite{Degiorgi1996PRL,Zhu2002PRB,Hu2008PRL}. The gap value $\Delta_{\text{CDW}}$ can be determined from the crossing point between $\sigma_{1}(\omega)$ at $T < T_{\text{CDW}}$ and that at $T$ slightly above $T_{\text{CDW}}$, such as 100~K. At 5~K, $\Delta_{\text{CDW}} = 86.5$~meV, which gives $2\Delta_{\text{CDW}}/k_{\text{B}}T_{\text{CDW}} \simeq 22$, a ratio much larger than the weak-coupling BCS value 3.53. This implies that the CDW in \CVS\ is unlikely to be driven by the conventional weak-coupling electron-phonon interaction~\cite{Gruner1988RMP}; a strong-coupling mechanism may be involved~\cite{Varma1983PRL}. The green solid circles in Fig.~\ref{CRefS1}(e) denotes $\Delta_{\text{CDW}}$ at different temperatures. In the proximity of \TCDW, $\Delta_{\text{CDW}}$ deviates from the BCS mean-field $T$ dependence (red solid curve) for $\Delta_0 = 86.5$~meV and $T_{\text{CDW}} = 91$~K, indicating an unconventional CDW in \CVS. This behavior is consistent with a recent study on \KVS~\cite{Uykur2021arXiv}. Furthermore, the residual Drude peak in the low-frequency range becomes extremely narrow, while the band at about 5500~\icm\ intensifies and shifts to higher frequency.

%
%

\begin{figure*}[tb]
\includegraphics[width=\textwidth]{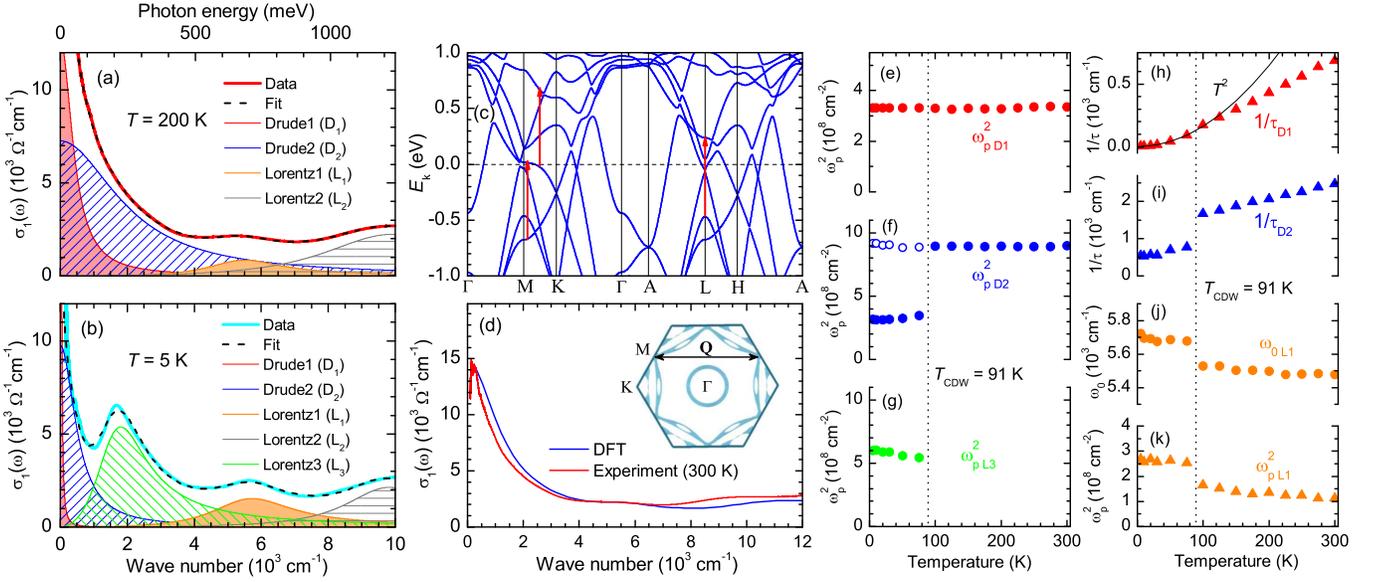}
\caption{(a) The red solid curve is $\sigma_{1}(\omega)$ of \CVS\ measured at 200~K. The black dashed line through the data represents the Drude-Lorentz fit, which is decomposed into a narrow Drude D$_{1}$ (red shaded area), a broad Drude D$_{2}$ (blue shaded area) and two Lorentz components L$_{1}$ and L$_{2}$, denoted by the orange and grey shaded regions, respectively. (b) The measured $\sigma_{1}(\omega)$ at 5~K (cyan solid curve) and the fit (black dashed curve). In addition to the components used for the fit at 200~K, an extra Lorentz component L$_{3}$ (green shaded area) is required to describe the opening of the CDW gap. (c) The calculated band structure for \CVS. (d) The calculated $\sigma_{1}(\omega)$ for \CVS\ (blue curve), which is compared to the measured $\sigma_{1}(\omega)$ at 300~K (red curve). Inset: The calculated Fermi surface of \CVS\ projected to the $k_x$-$k_y$ plane; the arrow denotes the nesting vector \textbf{Q}. (e)-(k) The $T$ dependence of the parameters extracted from the Drude-Lorentz fit.}
\label{CFit}
\end{figure*}

In order to quantitatively analyze the optical data, we fit the measured $\sigma_{1}(\omega)$ to the Drude-Lorentz model,
%
%
\begin{equation}
\sigma_{1}(\omega) = \frac{2\pi}{Z_{0}} \left[
   \sum_{k} \frac{\omega^{2}_{p,k}}{\tau_{k}(\omega^{2}+\tau_{k}^{-2})} +
   \sum_{j} \frac{\gamma_{j} \omega^{2} \omega_{p,j}^{2}}{(\omega_{0,j}^{2} - \omega^{2})^{2} + \gamma_{j}^{2} \omega^{2}}\right],
\label{DLModel}
\end{equation}
where $Z_{0} \simeq 377$~$\Omega$ is the impedance of free space. The first term is a sum of free-carrier Drude responses. Each is characterized by a plasma frequency $\omega_{p}$, with $\omega_{p}^{2}$ being proportional to the carrier density, and a scattering rate $1/\tau$ describing the width of the Drude profile at half maximum. In the second term, $\omega_{0,j}$, $\gamma_{j}$, and $\omega_{p,j}$ correspond to the resonance frequency, linewidth, and plasma frequency (strength) of the $j_{\text{th}}$ Lorentz oscillator, respectively. The red solid curve in Fig.~\ref{CFit}(a) is the measured $\sigma_{1}(\omega)$ at 200~K, and the dashed line through the data represents the fitting result, which is the superposition of a narrow Drude D$_{1}$ (red shaded area), a broad Drude D$_{2}$ (blue shaded area), and two Lorentz components L$_{1}$ and L$_{2}$ (the orange and grey shaded areas, respectively).

To ascertain the origin of the components in $\sigma_{1}(\omega)$, we calculated the band structure, the FS, and the $ab$-plane $\sigma_{1}(\omega)$ of \CVS\ within the density-functional-theory (DFT) framework implemented in the full-potential linearized augmented plane wave code WIEN2k~\cite{Blaha2001,Schwarz2003CMS,Abt1994PB,Ambrosch-Draxl2006CPC}. Figure~\ref{CFit}(c) displays the calculated band structure of \CVS, which is similar to previous calculations~\cite{Ortiz2019PRM,Ortiz2020PRL,Tan2021arXiv}. It is noteworthy that there are three types of bands crossing $E_{F}$: (i) a light electron-like band near the $\Gamma$ point, (ii) multiple Dirac bands with quasi-linear dispersion near the $K$ points, and (iii) heavy hole-like bands with weak dispersion which form saddle points near $E_{F}$ at $M$. The inset of Fig.~\ref{CFit}(d) shows the calculated FS of \CVS\ projected to the $k_x$-$k_y$ plane. The circle around the center of the Brillouin zone denotes a cylinder-like Fermi pocket produced by the light electron-like band near $\Gamma$; the multiple Dirac bands give rise to the FS sheets near $K$ (between $\Gamma$ and $K$); the FS near $M$ (between $M$ and $K$) are formed by the heavy hole-like bands with saddle points near $E_{\text{F}}$. Figure~\ref{CFit}(d) manifests that the calculated $\sigma_{1}(\omega)$ (blue solid curve) agrees quite well with the experimental $\sigma_{1}(\omega)$ (red solid curve).

Previous optical studies on Dirac or Weyl semimetals have shown that the Drude response associated with Dirac bands is very narrow~\cite{Xu2016PRB,Neubauer2016PRB,Schilling2017PRL,Xu2018PRL}, and the small scattering rate in these materials has been proved to stem from a protection mechanism that strongly suppresses backscattering~\cite{Liang2015NM,Liang2017PRL,Kumar2017NC}. A recent optical study on RhSi has corroborated that light electron bands give rise to a narrow Drude, while heavy hole bands produce a broad Drude~\cite{Ni2020NPJQM}. Furthermore, saddle points near $E_{\text{F}}$ act as scattering sinks, which lead to a large quasiparticle scattering rate~\cite{Rice1975PRL}. Based on these facts, it is reasonable to attribute D$_{1}$ to the light electron-like and multiple Dirac bands, and ascribe D$_{2}$ to the heavy hole bands having saddle points near $E_{F}$ at the $M$ points. L$_{1}$ and L$_{2}$ are associated with interband transitions. The energy of the interband transitions involving states near the saddle points at $M$ and $L$, as indicated by the red arrows in Fig.~\ref{CFit}(c), exactly matches the resonance frequency, namely the peak position ($\sim$0.7~eV), of L$_{1}$. This allows us to link L$_{1}$ to interband transitions involving the saddle points at $M$ and $L$. Here, we would like to point out that interband transitions between other bands may also contribute to L$_{1}$, but those involving the saddle points at $M$ and $L$ are expected to dominate, because near the saddle points, the density of states (DOS) is greatly enhanced due to the flatness of the bands.

Having elaborated the origin of each component, we examine $\sigma_{1}(\omega)$ in the CDW state and the evolution of the optical response with $T$. Figure~\ref{CFit}(b) depicts $\sigma_{1}(\omega)$ of \CVS\ at 5~K (cyan solid curve) and the fitting result (black dashed curve). Due to the opening of the CDW gap, D$_{2}$ is significantly suppressed, and a third Lorentz component L$_{3}$ (green shaded region) is introduced to describe the gap feature in $\sigma_{1}(\omega)$. The same approach has been used to describe the SDW gap in iron pnictides~\cite{Hu2008PRL,Nakajima2010PRB,Dai2016PRB}. Moreover, D$_{1}$ becomes very narrow to account for the extremely small $\rho(T)$ at 5~K; L$_{1}$ grows in spectral weight and shifts to higher frequency. In order to track the detailed $T$ dependence of the optical properties for \CVS, we apply the Drude-Lorentz fit to $\sigma_{1}(\omega)$ at all 15 measured temperatures, which returns the $T$ dependence of the parameters for all components. To achieve more accurate fits, the constraint condition $\sigma_{1}(\omega \rightarrow 0) \equiv 1/\rho(T)$ has been applied for the fit at all temperatures.

The $T$ dependence of $\omega^{2}_{p, D_1}$ and $\omega^{2}_{p, D_2}$ is traced out in Figs.~\ref{CFit}(e) and \ref{CFit}(f), respectively. Above \TCDW, neither $\omega^{2}_{p, D_1}$ nor $\omega^{2}_{p, D_2}$ exhibits observable $T$ dependence, indicating no change of the FSs. Upon entering the CDW state, while $\omega^{2}_{p, D_1}$ remains unchanged, a sharp drop in $\omega^{2}_{p, D_2}$ is observed. This suggests that the FSs associated with the light electron-like and Dirac bands are not affected by the CDW transition, whereas a large portion (about 65\%) of the FSs formed by the heavy hole bands with saddle points at $M$ is removed due to the formation of the CDW gap. The lost spectral weight from D$_{2}$ is transferred to higher energy. In order to locate the spectral weight, we plot $\omega^{2}_{p, L_3}$ as a function of $T$ in Fig.~\ref{CFit}(g). It is immediately obvious that the value of $\omega^{2}_{p, L_3}$ is approximately equal to the decrease of $\omega^{2}_{p, D_2}$, and $\omega^{2}_{p, L_3}$ exhibits the opposite $T$ dependence to $\omega^{2}_{p, D_2}$, implying that the lost spectral weight from D$_{2}$ is most likely transferred to L$_{3}$. To verify this assertion, $\omega^{2}_{p, D_2} + \omega^{2}_{p, L_3}$ is plotted as open circles in Fig.~\ref{CFit}(f), which reveals that $\omega^{2}_{p, D_2} + \omega^{2}_{p, L_3}$ in the CDW state has the same value as $\omega^{2}_{p, D_2}$ above \TCDW. This strongly suggests that the lost spectral weight from D$_{2}$ is fully captured by L$_{3}$, consistent with the expected behavior for a density-wave gap~\cite{Degiorgi1996PRL,Zhu2002PRB}. Moreover, the conservation of the spectral weight between D$_{2}$ and L$_{3}$ further confirms that the CDW transition bears no influence on $\omega^{2}_{p, D_1}$.

Figures~\ref{CFit}(h) and \ref{CFit}(i) display $1/\tau_{D_1}$ and $1/\tau_{D_2}$, respectively. Both diminishes linearly with decreasing $T$ above \TCDW. No abrupt change in $1/\tau_{D_1}$ is observed at \TCDW, again hinting that the light electron-like and Dirac bands are not involved in the CDW transition. In the low-temperature range, $1/\tau_{D_1}$ follows Fermi-liquid $T^{2}$ dependence as depicted by the solid curve in Fig.~\ref{CFit}(h), in accord with the expected behavior for bands with Dirac-like dispersion~\cite{Hosur2012PRL,Xu2016PRB}. In contrast, a sudden drop in $1/\tau_{D_2}$ occurs at \TCDW, suggesting that the quasiparticle scattering in the heavy hole bands with saddle points near $E_{F}$ is also altered by the formation of the CDW gap.

The evolution of L$_{1}$ is portrayed in Figs.~\ref{CFit}(j) and \ref{CFit}(k). Both the resonance frequency $\omega_{0, L_1}$ and the intensity $\omega^{2}_{p, L_1}$ of L$_{1}$ exhibit a jump at \TCDW. While the interband transitions at $L$ involve bands far away from $E_{\text{F}}$, which is unlikely to be affected by the CDW transition, the opening of the CDW gap in the heavy hole bands with saddle points near $E_{F}$ at $M$ can account for the behavior of L$_{1}$. Upon the CDW transition, the DOS near $E_{F}$ is depleted, leading to the formation of the CDW gap near $M$. The removed DOS is retrieved just above the gap energy, resulting in a pileup of DOS near the gap edge. This DOS pileup not only shifts the interband transitions towards higher energy, accounting for the jump in $\omega_{0, L_1}$, but also enhances their intensity due to the enhancement of DOS, which explains the increase of $\omega^{2}_{p, L_1}$.

%
Our experimental results provide important information for understanding the driving mechanism of the CDW order. Recent theoretical calculations~\cite{Tan2021arXiv} have suggested a 2$\times$2 CDW with an inverse star of David pattern as the ground state of \AVS, and also proposed that the CDW is induced by the Peierls instability related to the FS nesting and the softening of a breathing phonon mode of V atoms. The calculated phonon band structure shows a softening of acoustic phonon modes near the $M$ and $L$ points of the Brillouin Zone, indicating a strong instability. The $q$ vector of the soft mode at $M$ coincides with the nesting vector between neighboring saddle points at $M$. Consequently, the CDW instability significantly reduces the DOS at $E_{F}$ by suppressing the saddle points at $M$, which leads to the formation of the CDW gap in the heavy bands near the $M$ points. Assuming an electronically driven charge order and FS nesting effect, another theoretical work~\cite{Denner2021arXiv} has obtained a star of David CDW with orbital currents for \AVS, and the resulting band structure also exhibits a gap opening around the $M$ points. Although it is still controversial whether the FS nesting driving mechanism for a CDW in one-dimensional systems can be applied to two-dimensional materials~\cite{Johannes2008PRB}, Rice and Scott have demonstrated that in a two-dimensional system, the nesting of saddle points near $E_{\text{F}}$, which is not the usual FS nesting, is unstable against CDW formation~\cite{Rice1975PRL}.

\AVS\ have saddle points near $E_{\text{F}}$ at the $M$ points~\cite{Ortiz2019PRM,Ortiz2020PRL,Tan2021arXiv,Wang2021arXiv,Nakayama2021arXiv,Cho2021arXiv}, and these saddle points are connected by the nesting vector $\textbf{Q}$ as shown in the inset of Fig.~\ref{CFit}(d). Our experimental observations, i.e. the strong suppression of $\omega^{2}_{p, D_2}$, the shift of $\omega_{0, L1}$ to higher energy, as well as the increase of $\omega^{2}_{p, L1}$ upon the CDW transition, point to the formation of a CDW gap in the heavy bands having saddle points near $E_{F}$ at $M$. This has been confirmed by recent angle-resolved photoemission spectroscopy (ARPES) measurements~\cite{Wang2021arXiv,Nakayama2021arXiv,Cho2021arXiv}. The opening of the CDW gap in the heavy bands with saddle points at $M$ suggests that nesting of the saddle points at $M$ may play an important role in driving the CDW instability in \CVS.

%
%
To summarize, the optical properties of \CVS\ have been examined at numerous temperatures between 5 and 300~K. Above \TCDW, the optical conductivity can be well described by two Drude components: a narrow and a broad one. An investigation into the calculated band structure suggests that the narrow Drude is associated with a light electron-like and multiple Dirac bands, while the broad Drude arises from heavy hole bands with saddle points near $E_{F}$ at $M$. Below \TCDW, the opening of the CDW gap is clearly observed in $\sigma_{1}(\omega)$. The spectral weight of the broad Drude is substantially suppressed by the gap, while the narrow Drude remains unchanged. In addition, interband transitions that involve the electronic states near the saddle points at $M$ gain some weight and shift to higher energy. These observations signify that the CDW instability in \CVS\ is intimately related to the nesting of the saddle points at $M$.

%
%

\begin{acknowledgments}
We thank S.~L. Yu, Q.~H. Wang, R. Thomale, B.~H. Yan, H. Miao, B. Xu, R. Yang, and J.~B. Qi for helpful discussions. We gratefully acknowledge financial support from the National Key R\&D Program of China (Grants No. 2016YFA0300401 and 2020YFA0308800), the National Natural Science Foundation of China (Grants No. 11874206, 12061131001, 92065109, 11734003 and 11904294), the Fundamental Research Funds for the Central Universities (Grant No. 020414380095), Jiangsu shuangchuang program, the Beijing Natural Science Foundation (Grant No. Z190006), and the Beijing Institute of Technology Research Fund Program for Young Scholars (Grant No. 3180012222011).
\end{acknowledgments}

%

\end{document}